\documentstyle[
  aps,prl,epsf,multicol,amsmath,graphics,psfrag,graphicx,pstcol,psfig
]{revtex}

\begin{document}
\draft

\title{
  Renormalization of the electron--phonon interaction: a reformulation of the 
  BCS--gap equation 
}

\author{A.~H\"{u}bsch and K.~W.~Becker}
\address{
  Institut f\"{u}r Theoretische Physik,
  Technische Universit\"{a}t Dresden, D-01062 Dresden, Germany
}

\date{\today}
\maketitle

\begin{abstract}
A recently developed renormalization approach is used to study the 
electron-phonon coupling in many-electron systems. By starting from  an 
Hamiltonian which includes a small gauge symmetry breaking field, we 
directly derive a BCS-like equation for the energy gap from  
the renormalization approach. The effective  
electron-electron interaction for Cooper pairs does not contain 
any singularities. 
Furthermore, it is found that phonon-induced particle-hole excitations only 
contribute to the attractive electron-electron interaction if their energy
difference is smaller than the phonon energy. 
\end{abstract}


\widetext

%%%%%%%%%%%%%%%%%%%%%%%%%%%%%%%%%%%%%%%%%%%%%%%%%%%%%%%%%%%%%%%%%%%%%%%%%%%%%%%
\section{Introduction}

The famous BCS-theory \cite{BCS} of superconductivity is essentially based on 
the analysis of attractive interactions between electrons of many-particle 
systems \cite{Cooper}. As was pointed out 
by Fr\"{o}hlich \cite{Froehlich} such an interaction can result 
from an effective coupling between electrons mediated via phonons. 
The recent discovery of superconductivity  
in magnesium diboride MgB$_{2}$ \cite{Nagamatsu} below a rather high 
$T_c$ of about 39 K has attracted again 
a lot of interest on this classical 
scenario of a phonon-mediated superconductivity. 
However, the electron-electron interaction 
derived by Fr\"{o}hlich \cite{Froehlich} contains some problems. There are 
certain regions in momentum space where the attractive interaction 
becomes singular and changes its sign due to a vanishing energy 
denominator. 

Recently, effective phonon-induced electron-electron interactions were 
also derived \cite{Lenz,Mielke} by use of Wegner's flow equation 
method \cite{Wegner} and by a  similarity renormalization proposed by 
Glatzek and Wilson \cite{Wilson1,Wilson2}. The main idea of these 
approaches is to perform a continuous unitary transformation which  
leads to an expression for an
effective electron-electron 
interaction which is less singular than Fr\"{o}hlich's result \cite{Froehlich}.

Recently, we have developed a renormalization approach which is based on 
perturbation theory \cite{Becker}. This approach resembles 
Wegner's flow equation method \cite{Wegner} and the similarity renormalization 
\cite{Wilson1,Wilson2} in some aspects. Therefore, the investigation of an 
effective phonon-induced electron-electron interaction is very useful to 
compare the three methods in more details. Therefore, in this paper we 
directly diagonalize the classical problem of interacting electrons and 
phonons by use of the new renormalization technique \cite{Becker}. the
Hamiltonian is given by 
\begin{eqnarray}
  \label{1}
  {\cal H} &=& 
  \sum_{{\bf k},\sigma} \varepsilon_{\bf k} \,
    c_{{\bf k},\sigma}^{\dagger}c_{{\bf k},\sigma}  +
  \sum_{\bf q} \omega_{\bf q} \,
    b_{\bf q}^{\dagger}b_{\bf q} +
  \sum_{{\bf k},{\bf q},\sigma}
  \left(
    g_{\bf q} \, 
    c_{{\bf k},\sigma}^{\dagger} c_{({\bf k}+{\bf q}),\sigma} 
    b_{\bf q}^{\dagger} + 
    g_{\bf q}^{*} \,
    c_{({\bf k}+{\bf q}),\sigma}^{\dagger} c_{{\bf k},\sigma}
    b_{\bf q}
  \right),
\end{eqnarray}
which will be used to describe superconducting properties. 
In \eqref{1}
$c_{{\bf k},\sigma}^{\dagger}$ and $c_{{\bf k},\sigma}$ are the 
usual creation and 
annihilation operators for electrons with wave vector ${\bf k}$ and spin 
$\sigma$.  $b_{\bf q}^{\dagger}$ and $b_{\bf q}$ denote phonon operators
with phonon energies    $\omega_{\bf q}$.  
The electron excitation energies $\varepsilon_{\bf k}$ 
are measured from the chemical potential $\mu$.

The paper is organized as follows. In the next section we briefly 
repeat our recently developed renormalization approach \cite{Becker}. 
In 
Sec.~\ref{El-Ph} this approach will be  
applied to the electron-phonon system \eqref{1} in order
to derive a BCS-like  gap equation. 
Furthermore, effective electron-electron interaction 
derived in this framework will be  compared 
with the results from former approaches 
\onlinecite{Froehlich,Lenz,Mielke}. Finally, our 
conclusions are presented in Sec.~\ref{Conc}.

%%%%%%%%%%%%%%%%%%%%%%%%%%%%%%%%%%%%%%%%%%%%%%%%%%%%%%%%%%%%%%%%%%%%%%%%%%%%%%%
\section{Projector-based renormalization method (PRM)}\label{Ren}

The PRM \cite{Becker} starts from the  decomposition of a given 
many-particle Hamiltonian ${\cal H}$ into an unperturbed part ${\cal H}_{0}$ 
and into a perturbation ${\cal H}_{1}$
\begin{eqnarray}
  \label{2}
  {\cal H} &=& {\cal H}_{0} + \varepsilon {\cal H}_{1} =: H(\varepsilon). 
\end{eqnarray}
We assume that the eigenvalue problem 
${\cal H}_{0} |n\rangle = E^{(0)}_n |n\rangle$ of the unperturbed part 
${\cal H}_{0}$ is known. The parameter $\varepsilon$ accounts for the order of 
perturbation processes discussed below. Let us define projection operators 
${\bf P}_{\lambda}$ and ${\bf Q}_{\lambda}$ by
\begin{eqnarray}
  \label{3}
  {\bf P}_{\lambda}A &=& \sum_{|E_n^{(0)} - E_m^{(0)}| \leq \lambda}
  |n\rangle \langle m| \, \langle n| A  |m\rangle \qquad\mbox{and} \\
  \label{4}
  {\bf Q}_{\lambda} &=& {\bf 1} - {\bf P}_{\lambda}.
\end{eqnarray}
${\bf P}_{\lambda}$ and ${\bf Q}_{\lambda}$ are super-operators acting 
on usual operators $A$ of the unitary space. Here,
${\bf P}_{\lambda}$ projects on 
that part of $A$ which is formed by all transition operators 
 $|n\rangle \langle m|$ with 
energy differences $|E_n^{(0)} - E_m^{(0)}|$ less or equal to a given cutoff 
$\lambda$. The cutoff $\lambda$ is  smaller than the cutoff $\Lambda$ of the 
original Hamiltonian ${\cal H}$.  
${\bf Q}_{\lambda}$ is orthogonal to ${\bf P}_{\lambda}$ 
and projects on high energy transitions larger than $\lambda$.

The aim is to transform the initial 
Hamiltonian ${\cal H}$ into an 
effective Hamiltonian ${\cal H}_{\lambda}$ which  
has no matrix elements between  eigenstates of ${\cal H}_0$ 
with energy differences larger than $\lambda$. 
 ${\cal H}_{\lambda}$ will be  constructed 
by use of an unitary transformation
\begin{eqnarray}
  \label{5}
  {\cal H}_{\lambda} &=&
  e^{X_{\lambda}} \, {\cal H} \, e^{-X_{\lambda}} .
\end{eqnarray}
Due to construction the effective Hamiltonian 
${\cal H}_{\lambda}$ will therefore have the same
eigenspectrum as the original Hamiltonian ${\cal H}$. The generator 
$X_{\lambda}$ of the transformation is anti-Hermitian, 
$X^{\dagger}_{\lambda}=-X_{\lambda}$. To find an appropriate generator 
$X_{\lambda}$  we use the condition that ${\cal H}_{\lambda}$
has no matrix elements with transition energies larger than $\lambda$, i.e., 
\begin{eqnarray}
  \label{6}
  {\bf Q}_{\lambda}{\cal H}_{\lambda} &=& 0
\end{eqnarray}
has to be fulfilled. By assuming that $X_{\lambda}$ 
can be written as a power series in the perturbation parameter $\varepsilon$
\begin{eqnarray}
  \label{7}
  X_{\lambda} &=&
    \varepsilon X_{\lambda}^{(1)} + \varepsilon^{2} X_{\lambda}^{(2)} +
    \varepsilon^{3} X_{\lambda}^{(3)} + \dots
\end{eqnarray}
the effective Hamiltonian ${\cal H}_{\lambda}$ can be expanded in a
power series in $\varepsilon$ as well
\begin{eqnarray}
  \label{8}
  {\cal H}_{\lambda} &=&
  {\cal H}_{0} +
  \varepsilon
  \left\{
    {\cal H}_{1} +
    \left[
      X_{\lambda}^{(1)}, {\cal H}_{0}
    \right]
  \right\} + \\
  &&
  + \varepsilon^{2}
  \left\{
    \left[
      X_{\lambda}^{(1)}, {\cal H}_{1}
    \right] +
    \left[
      X_{\lambda}^{(2)}, {\cal H}_{0}
    \right] +
    \frac{1}{2!}
    \left[
      X_{\lambda}^{(1)},
      \left[
        X_{\lambda}^{(1)}, {\cal H}_{0}
      \right]
    \right]
  \right\} + {\cal O}(\varepsilon^{3}) .\nonumber
\end{eqnarray}
Now, the high-energy parts parts  ${\bf Q}_{\lambda} X_{\lambda}^{(n)}$ 
of   $X_{\lambda}^{(n)}$ can successively be
determined from Eq.~\eqref{6} whereas the low-energy parts 
${\bf P}_{\lambda} X_{\lambda}^{(n)}$ can still be chosen arbitrarily. In the 
following we use for convenience
$
  {\bf P}_{\lambda} X_{\lambda}^{(1)} =
  {\bf P}_{\lambda} X_{\lambda}^{(2)} = 0
$
so that for the effective Hamiltonian ${\cal H}_{\lambda}$ 
up to second order in ${\cal H}_1$ follows
\begin{eqnarray}
  \label{9}
  {\cal H}_{\lambda} &=&
  {\cal H}_{0} + {\bf P}_{\lambda}{\cal H}_{1} -
  \frac{1}{2} {\bf P}_{\lambda}
  \left[
    ( {\bf Q}_{\lambda}{\cal H}_{1}), \frac{1}{{\bf L}_{0}}
    ( {\bf Q}_{\lambda}{\cal H}_{1})
  \right] - 
  {\bf P}_{\lambda}
  \left[
    ({\bf P}_{\lambda}{\cal H}_{1}), \frac{1}{ {\bf L}_{0} }
    ({\bf Q}_{\lambda}{\cal H}_{1})
  \right]
\end{eqnarray}
Here $\varepsilon$ was set equal to $1$. The quantity ${\bf L}_{0}$ 
in \eqref{9} denotes the 
Liouville operator of the unperturbed Hamiltonian. It is defined by 
${\bf L}_{0}=[{\cal H}_{0},A]$ for any operator $A$. Note that the 
perturbation expansion \eqref{9} can easily be extended to higher orders in 
$\varepsilon$. One should also note that 
the correct size dependence of the Hamiltonian 
is automatically guaranteed due to the commutators 
appearing in \eqref{9}.

Next, let us use this perturbation theory to establish a renormalization 
approach by successively reducing the cutoff energy $\lambda$. In particular, 
instead of eliminating high-energy excitations in one step a 
sequence of stepwise transformations is used. Thereby, 
we obtain an effective
model which becomes diagonal in the limit $\lambda\rightarrow 0$. In an 
infinitesimal formulation, the method yields renormalization equations as 
function of the cutoff $\lambda$. To find these equations we  
start from the renormalized Hamiltonian
\begin{eqnarray}
  \label{10}
  {\cal H}_{\lambda} &=& {\cal H}_{0,\lambda} + {\cal H}_{1,\lambda}
\end{eqnarray}
after all excitations with energy differences larger than $\lambda$ have been
eliminated. Now we perform an additional renormalization of 
${\cal H}_{\lambda}$ by eliminating all excitations inside an  energy shell 
between $\lambda$ and a smaller energy cutoff $(\lambda-\Delta\lambda)$ 
where $\Delta\lambda>0$. The new Hamiltonian is found by use of \eqref{9}
\begin{eqnarray}
  \label{11}
  {\cal H}_{(\lambda-\Delta\lambda)} &=&
  {\bf P}_{(\lambda-\Delta\lambda)} {\cal H}_{\lambda} - 
  \frac{1}{2} {\bf P}_{(\lambda-\Delta\lambda)}
  \left[
    ( {\bf Q}_{(\lambda-\Delta\lambda)}{\cal H}_{1,\lambda}),
    \frac{1}{{\bf L}_{0,\lambda}}
    ( {\bf Q}_{(\lambda-\Delta\lambda)}{\cal H}_{1,\lambda})
  \right] +  \\
  &&  -
  {\bf P}_{(\lambda-\Delta\lambda)}
  \left[
    ({\bf P}_{(\lambda-\Delta\lambda)}{\cal H}_{1,\lambda}),
    \frac{1}{ {\bf L}_{0,\lambda} }
    ({\bf Q}_{(\lambda-\Delta\lambda)}{\cal H}_{1,\lambda})
  \right] . \nonumber
\end{eqnarray}
Here, ${\bf L}_{0,\lambda}$ denotes the Liouville operator with respect to the 
unperturbed part ${\cal H}_{0,\lambda}$ of the $\lambda$ dependent
Hamiltonian ${\cal H}_{\lambda}$. Note that the flow equations derived from 
Eq.~\eqref{11} will lead to an approximative renormalization of 
${\cal H}_{\lambda}$ because only contributions up to second order in 
${\cal H}_{1,\lambda}$ are included in Eq. \eqref{11}.
For a concrete evaluation of 
Eq.~\eqref{11} it is useful to divide the second order 
term on the r.h.s into two parts: The first one connects 
eigenstates of ${\cal H}_{0,\lambda}$ with 
the same energy. This part commutes with 
${\cal H}_{0,\lambda}$ and can therefore be considered 
as renormalization of the unperturbed Hamiltonian
\begin{eqnarray}
  \label{12}
  {\cal H}_{0,(\lambda-\Delta\lambda)} - {\cal H}_{0,\lambda} &=&
  - \frac{1}{2} \, {\bf P}_{0}
  \left[
    ( {\bf Q}_{(\lambda-\Delta\lambda)}{\cal H}_{1,\lambda}),
    \frac{1}{{\bf L}_{0,\lambda}}
    ( {\bf Q}_{(\lambda-\Delta\lambda)}{\cal H}_{1,\lambda})
  \right] .
\end{eqnarray}
In contrast the second part connects eigenstates of ${\cal H}_{0,\lambda}$
with different energies and represents a renormalization of the interaction 
part of the Hamiltonian
\begin{eqnarray}
  \label{13}
  {\cal H}_{1,(\lambda-\Delta\lambda)} -
  {\bf P}_{(\lambda-\Delta\lambda)}{\cal H}_{1,\lambda}
  &=&  - {\bf P}_{(\lambda-\Delta\lambda)}
  \left[
    ({\bf P}_{(\lambda-\Delta\lambda)}{\cal H}_{1,\lambda}),
    \frac{1}{ {\bf L}_{0,\lambda} }
    ({\bf Q}_{(\lambda-\Delta\lambda)}{\cal H}_{1,\lambda})
  \right]  \\
  && - \frac{1}{2}
  \left(
    {\bf P}_{(\lambda-\Delta\lambda)} - {\bf P}_{0}
  \right)
  \left[
    ( {\bf Q}_{(\lambda-\Delta\lambda)}{\cal H}_{1,\lambda}),
    \frac{1}{{\bf L}_{0,\lambda}}
    ( {\bf Q}_{(\lambda-\Delta\lambda)}{\cal H}_{1,\lambda})
  \right].  \nonumber
\end{eqnarray}
Note that for small $\Delta\lambda$  only the mixed term, i.e., the 
first part on the r.h.s.~of Eq.~\eqref{13},  contributes to the 
renormalization of ${\cal H}_{1,\lambda}$
\begin{eqnarray}
  \label{14}
  {\cal H}_{1,(\lambda-\Delta\lambda)} -
  {\bf P}_{(\lambda-\Delta\lambda)}{\cal H}_{1,\lambda}
  &\approx&
  - {\bf P}_{(\lambda-\Delta\lambda)}
  \left[
    ({\bf P}_{(\lambda-\Delta\lambda)}{\cal H}_{1,\lambda}),
    \frac{1}{ {\bf L}_{0,\lambda} }
    ({\bf Q}_{(\lambda-\Delta\lambda)}{\cal H}_{1,\lambda})
  \right] .
\end{eqnarray}
In the limit $\Delta\lambda \rightarrow 0$, i.e.~for vanishing shell width, 
equations \eqref{12} and \eqref{14} lead to differential 
equations for the Hamiltonian as function of the cutoff energy 
$\lambda$. The resulting equations for the parameters of the Hamiltonian are 
called flow equations. Their solution depend on the initial values of the 
parameters of the Hamiltonian. Note that for $\lambda \rightarrow 0$ the 
resulting Hamiltonian only consists of the unperturbed part 
${\cal H}_{(\lambda \rightarrow 0)}$ so that an effectively diagonal 
Hamiltonian is obtained.

%%%%%%%%%%%%%%%%%%%%%%%%%%%%%%%%%%%%%%%%%%%%%%%%%%%%%%%%%%%%%%%%%%%%%%%%%%%%%%%
\section{Application to the electron-phonon system}\label{El-Ph}

In this section we apply the renormalization approach discussed above to the 
system \eqref{1} of interacting electrons and phonons. The aim is 
to decouple the electron and the phonon subsystems.  
The Hamiltonian \eqref{1} is gauge invariant. In 
contrast, a BCS-like Hamiltonian breaks this symmetry. 
Thus, in order to describe 
the superconducting state of the system, the 
renormalized Hamiltonian should contain a symmetry breaking field. 
Therefore, our starting Hamiltonian ${\cal H}_{\lambda}$ reads 
\begin{eqnarray}
  \label{15}
  {\cal H}_{\lambda} &=& {\cal H}_{0,\lambda} + {\cal H}_{1,\lambda}, 
\end{eqnarray}
after all excitations with energies larger than $\lambda$ have  been 
eliminated, where 
\begin{eqnarray}
\label{16}
  {\cal H}_{0,\lambda} &=&
  \sum_{{\bf k},\sigma} \varepsilon_{\bf k} \,
    c_{{\bf k},\sigma}^{\dagger}c_{{\bf k},\sigma}  +
  \sum_{\bf q} \omega_{\bf q} \,
    b_{\bf q}^{\dagger}b_{\bf q} - 
  \sum_{\bf k}
  \left(
    \Delta_{{\bf k},\lambda} \,
    c_{{\bf k},\uparrow}^{\dagger} c_{-{\bf k},\downarrow}^{\dagger} + 
    \Delta_{{\bf k},\lambda}^{*} \,
    c_{-{\bf k},\downarrow} c_{{\bf k},\uparrow} 
  \right) + 
  C_{\lambda},  \\
 && \nonumber \\
\label{17}  
{\cal H}_{1,\lambda} &=& 
  {\bf P}_{\lambda} {\cal H}_{1} \,=\,
  {\bf P}_{\lambda}
  \sum_{{\bf k},{\bf q},\sigma}
  \left(
    g_{\bf q} \, 
    c_{{\bf k},\sigma}^{\dagger} c_{({\bf k}+{\bf q}),\sigma} 
    b_{\bf q}^{\dagger} + 
    g_{\bf q}^{*} \,
    c_{({\bf k}+{\bf q}),\sigma}^{\dagger} c_{{\bf k},\sigma}
    b_{\bf q}
  \right) .
\end{eqnarray}
The 'fields' $\Delta_{{\bf k},\lambda}$ and 
$\Delta_{{\bf k},\lambda}^*$ in 
 ${\cal H}_{0,\lambda}$ couple to  the operators  
$c_{{\bf k},\uparrow}^{\dagger} c_{-{\bf k},\downarrow}^{\dagger}$
and  $c_{-{\bf k},\downarrow} c_{{\bf k},\uparrow}$ and 
break the gauge invariance.  They will take over the role of the 
superconducting gap function but still depend on $\lambda$.
The initial values for  $\Delta_{{\bf k}, \lambda}$ and 
the energy shift $C_{\lambda}$
are those of  the original model
\begin{eqnarray}
  \label{18}
  \Delta_{{\bf k},(\lambda=\Lambda)}  &=& 0, \quad 
  C_{(\lambda=\Lambda)} \,=\,0.
\end{eqnarray}
Note that renormalization contributions to the 
electron energies $\varepsilon_{\bf k}$, the phonon energies 
$\omega_{\bf q}$, and the electron-phonon interactions $g_{\bf q}$ 
 have been neglected in \eqref{15}.
Also, additional interactions which would appear due to  
renormalization processes have been omitted.
Let us first solve  the eigenvalue problem of 
${\cal H}_{0,\lambda}$. For this purpose, we perform a 
Bogoliubov transformation \cite{Bogoliubov} and introduce new $\lambda$ 
dependent fermionic quasi-particles
\begin{eqnarray}
  \label{19}
  \alpha_{{\bf k},\lambda}^{\dagger} &=& 
  u_{{\bf k},\lambda}^{*} c_{{\bf k},\uparrow}^{\dagger} - 
  v_{{\bf k},\lambda}^{*} c_{-{\bf k},\downarrow} ,\\
  \beta_{{\bf k},\lambda}^{\dagger} &=& 
  u_{{\bf k},\lambda}^{*} c_{-{\bf k},\downarrow}^{\dagger} + 
  v_{{\bf k},\lambda}^{*} c_{{\bf k},\uparrow}\nonumber
\end{eqnarray}
where 
\begin{eqnarray}
  \label{20}
  \left|
    u_{{\bf k},\lambda}
  \right|^{2}
  &=&
  \frac{1}{2}
  \left(
    1 + 
    \frac{
      \varepsilon_{\bf k}
    }{
      \sqrt{
        \varepsilon_{\bf k}^{2} + 
        \left| \Delta_{{\bf k},\lambda}  \right|^{2}
      }
    }
  \right), \\
  \left| v_{{\bf k},\lambda} \right|^{2}
  &=&
  \frac{1}{2}
  \left(
    1 - 
    \frac{
      \varepsilon_{\bf k}
    }{
      \sqrt{
        \varepsilon_{\bf k}^{2} + 
        \left| \Delta_{{\bf k},\lambda}  \right|^{2}
      }
    }
  \right). \nonumber
\end{eqnarray}
${\cal H}_{0,\lambda}$ can be rewritten as
\begin{eqnarray}
  \label{21}
  {\cal H}_{0,\lambda} &=& 
  \sum_{\bf k} E_{{\bf k},\lambda}
  \left(
    \alpha_{{\bf k},\lambda}^{\dagger} \alpha_{{\bf k},\lambda} +
    \beta_{{\bf k},\lambda}^{\dagger} \beta_{{\bf k},\lambda}
  \right) + 
  \sum_{\bf k}
  \left(
    \varepsilon_{\bf k} - E_{{\bf k},\lambda}
  \right) + 
  \sum_{\bf q} \omega_{\bf q} \, b_{\bf q}^{\dagger}b_{\bf q} + C_{\lambda} 
\end{eqnarray}
where the fermionic excitation energies are given by  
$
  E_{{\bf k},\lambda} = 
  \sqrt{
    \varepsilon_{\bf k}^{2} + \left| \Delta_{{\bf k},\lambda}  \right|^{2}
  }
$.
\bigskip

Next let us  eliminate all excitations within the energy shell between 
$\lambda$ and $(\lambda-\Delta\lambda)$  
by applying the renormalization scheme of section II. 
We are primarily interested in the renormalization of the 
gap function $\Delta_{{\bf k}, \lambda}$. Note that for this case we have 
to consider the renormalization contribution given by \eqref{14}
\begin{eqnarray}
  \label{22}
  \delta{\cal H}_1(\lambda,\Delta\lambda) &:=& 
  {\cal H}_{1,(\lambda-\Delta\lambda)} - 
  {\bf P}_{(\lambda-\Delta\lambda)} {\cal H}_{1,\lambda}.
\end{eqnarray}
The reason is that the renormalization \eqref{12} of 
${\cal H}_{0, \lambda}$ only gives contributions which connects 
eigenstates of  ${\cal H}_{0,\lambda}$ with the same energy. Thus, this
part only changes the quasiparticle 
energies from $ E_{{\bf k},\lambda}$ to  
 $ E_{{\bf k},(\lambda -\Delta \lambda)}$. 
In contrast, the renormalization
\eqref{22} changes the relative weight of the operator 
terms in \eqref{16} and is exactly the renormalization needed 
to describe the flow of $\Delta_{{\bf k},\lambda}$ which will be
discussed in the following.

There is no principle problem to evaluate the renormalization 
contributions \eqref{14}. First, one expresses
the creation and annihilation operators by the quasiparticle
operators \eqref{19} and uses the relation 
$
  {\bf L}_{0,\lambda} \alpha_{k,\lambda}^{\dagger}=
  E_{{\bf k},\lambda} \alpha_{{\bf k}, \lambda}^{\dagger}
$
and an equivalent relation for $\beta_{{\bf k}, \lambda}^{\dagger}$ to 
evaluate the denominator in Eq. \eqref{14}. Then the quasi-particle 
operators \eqref{18} have to be transformed back to the original 
electron operators. The main reason for this procedure is the fact 
that the Bogoliubov transformation \eqref{19},\eqref{20} depends on the cutoff 
$\lambda$. Thereby, a lot of terms arise which contribute to the  
renormalization \eqref{14}. For convenience, we evaluate the denominator in 
Eq. \eqref{14} by use of the assumption 
$\varepsilon_{\bf k}^{2} \gg |\Delta_{{\bf k},\lambda}|^{2}$. As it turns out, 
the resulting flow equations still contain sums over ${\bf k}$. Note that the 
approximation used is valid for most of the ${\bf k}$ dependent terms (except 
for those ${\bf k}$ values close to the Fermi momentum). Thus, it does not 
strongly affected the renormalization contributions. The two operator 
expressions contributing to the commutator in \eqref{14} are
given in this approximation by 
\begin{eqnarray}
  \label{23}
  {\bf P}_{(\lambda-\Delta\lambda)} {\cal H}_{1,\lambda}
  &=&
  \sum_{{\bf k},{\bf q},\sigma}
  \Theta\left[
    (\lambda-\Delta\lambda) -
    \left| 
      \varepsilon_{\bf k} - \varepsilon_{({\bf k}+{\bf q})} + 
      \omega_{\bf q}
    \right|
  \right]
  \left\{
    g_{\bf q} \, 
    c_{{\bf k},\sigma}^{\dagger} c_{({\bf k}+{\bf q}),\sigma} 
    b_{\bf q}^{\dagger}
    + {\rm h.c.}
  \right\},\\
&& \nonumber \\
  \frac{1}{{\bf L}_{0,\lambda}} 
  {\bf Q}_{(\lambda-\Delta\lambda)} {\cal H}_{1,\lambda}
  &=& 
  \label{24}
  \sum_{{\bf k},{\bf q},\sigma}
  \frac{
    \delta\Theta_{{\bf k},{\bf q}}(\lambda,\Delta\lambda)
  }{
    \varepsilon_{\bf k} - \varepsilon_{({\bf k}+{\bf q})} + \omega_{\bf q}
  }
  \left\{
        g_{\bf q} \, 
    c_{{\bf k},\sigma}^{\dagger} c_{({\bf k}+{\bf q}),\sigma} 
    b_{\bf q}^{\dagger}
    - {\rm h.c.}
  \right\} \\
&& \nonumber
\end{eqnarray}
where
\begin{eqnarray}
  \label{25}
  \delta\Theta_{{\bf k},{\bf q}}(\lambda,\Delta\lambda) &=& 
  \Theta\left[
    \left| 
      \varepsilon_{\bf k} - \varepsilon_{({\bf k}+{\bf q})} + 
      \omega_{\bf q}
    \right| - (\lambda-\Delta\lambda)
  \right] -
    \Theta\left[
    \left| 
      \varepsilon_{\bf k} - \varepsilon_{({\bf k}+{\bf q})} + 
      \omega_{\bf q}
    \right| - \lambda
  \right] 
\end{eqnarray}
describes the restriction to excitations on the energy shell 
$\Delta \lambda$. We are not interested in the 
renormalization of the phonon modes. Therefore all contributions 
including phonon operators are neglected. By using 
\eqref{23} and \eqref{24} we then find from \eqref{14} 
\begin{eqnarray}
  &&\nonumber\\
  \delta{\cal H}_1(\lambda,\Delta\lambda) &=&
  - {\bf P}_{(\lambda-\Delta\lambda)}
  \sum_{{\bf k},{\bf k}^{\prime},{\bf q},\sigma,\sigma^{\prime}}
  \frac{
    \left| g_{\bf q} \right|^{2} 
    \delta\Theta_{{\bf k},{\bf q}}(\lambda,\Delta\lambda)
    \Theta\left[
      (\lambda-\Delta\lambda) -
      \left| 
        \varepsilon_{{\bf k}^{\prime}} - 
        \varepsilon_{({\bf k}^{\prime}+{\bf q})} + \omega_{\bf q}
      \right|
    \right]
  }{
    \varepsilon_{\bf k} - \varepsilon_{({\bf k}+{\bf q})} + \omega_{\bf q}
  } \times
  \nonumber\\
  &&
  \label{26}
  \qquad\qquad\qquad
  \times\left\{
    c_{({\bf k}+{\bf q}),\sigma}^{\dagger} c_{{\bf k},\sigma} 
    c_{{\bf k}^{\prime},\sigma^{\prime}}^{\dagger}
    c_{({\bf k}^{\prime}+{\bf q}),\sigma^{\prime}} 
    + {\rm h.c}
  \right\}.
\end{eqnarray}

In the following we restrict ourselves to renormalization 
contributions which lead to the formation of Cooper pairs. 
Consequently, the conditions ${\bf k}^{\prime}=-({\bf k}+{\bf q})$ and 
$\sigma^{\prime}=-\sigma$ have to be fulfilled so that
\begin{eqnarray}
  && \nonumber\\[-4ex]
  \label{27}
&&  - \lim_{\Delta\lambda\rightarrow 0}
  \frac{\delta{\cal H}_1(\lambda,\Delta\lambda)}{\Delta\lambda} = \\[1.5ex]
 && \hspace*{0.5cm}  =
  \sum_{{\bf k},{\bf q},\sigma}
  \Theta\left[
    \lambda - 
    2\left|
      \varepsilon_{\bf k} - \varepsilon_{({\bf k}+{\bf q})}
    \right|
  \right]
  \Theta\left[
    \lambda -
    \left| 
      \varepsilon_{({\bf k}+{\bf q})} - \varepsilon_{\bf k} + \omega_{\bf q}
    \right|
  \right]
  \delta\left(
    \left| 
      \varepsilon_{\bf k} - \varepsilon_{({\bf k}+{\bf q})} + \omega_{\bf q}
    \right|
    - \lambda
  \right) \times
  \nonumber\\
  &&
  \qquad\qquad
  \times \frac{
    \left| g_{\bf q} \right|^{2}
  }{
    \varepsilon_{\bf k} - \varepsilon_{({\bf k}+{\bf q})} + \omega_{\bf q}
  }
  \left\{
    c_{({\bf k}+{\bf q}),\sigma}^{\dagger} 
    c_{-({\bf k}+{\bf q}),-\sigma}^{\dagger}
    c_{-{\bf k},-\sigma} 
    c_{{\bf k},\sigma} 
    + {\rm h.c}
  \right\}\nonumber \\[-2ex]
&& \nonumber 
\end{eqnarray}
results from \eqref{26}. Here, 
$
  \Theta\left[
    \lambda - 
    2\left|
      \varepsilon_{\bf k} - \varepsilon_{({\bf k}+{\bf q})}
    \right|
  \right]
$
is due to the projector operator ${\bf P}_{(\lambda-\Delta\lambda)}$ in 
\eqref{26}. Note that the differential expression on the l.h.s.~of
\eqref{27} is different from  
the differential $d{\cal H}_{1,\lambda}/ d\lambda$. This follows 
from the definition of 
$
  \delta {\cal H}_{1}(\lambda, \Delta \lambda) = 
  {\cal H}_{1,(\lambda -\Delta \lambda)} - 
  {\bf P}_{(\lambda -\Delta \lambda)}{\cal H}_{1, \lambda}
$ 
which differs from  
$
  \Delta {\cal H}_{1}(\lambda, \Delta \lambda) =
  {\cal H}_{1,(\lambda -\Delta \lambda)} - 
  {\cal H}_{1, \lambda}
$ 
due to the second term. Next we can 
simplify the $\Theta$-functions in Eq.~\eqref{27} by discussing  
$\varepsilon_{\bf k}\geq \varepsilon_{({\bf k}+{\bf q})}$ and 
$\varepsilon_{({\bf k}+{\bf q})} > \varepsilon_{\bf k}$ separately.
There are no contributions from the latter case. For 
$\varepsilon_{\bf k}\geq \varepsilon_{({\bf k}+{\bf q})}$ the 
contribution from the first term in the curly bracket in
\eqref{27} and from its
conjugate can be combined. By exploiting the $\Theta$-functions the
result can be rewritten as 
\begin{eqnarray}
  \label{28}
  - \lim_{\Delta\lambda\rightarrow 0}
  \frac{\delta{\cal H}_1(\lambda,\Delta\lambda)}{\Delta\lambda} 
&=&
  \sum_{{\bf k},{\bf q},\sigma} 
  \delta\left(
    \left| 
      \varepsilon_{\bf k} - \varepsilon_{({\bf k}+{\bf q})}
    \right|
    + \omega_{\bf q} - \lambda
  \right) 
  \frac{
    \left| g_{\bf q} \right|^{2}
    \Theta\left[
      \omega_{\bf q} - 
      \left| \varepsilon_{\bf k} - \varepsilon_{({\bf k}+{\bf q})} \right|
    \right]
  }{
    \left|
      \varepsilon_{\bf k} - \varepsilon_{({\bf k}+{\bf q})}
    \right| + \omega_{\bf q}
  }\times \nonumber \\[1ex]
&&
  \qquad\qquad\times
  c_{({\bf k}+{\bf q}),\sigma}^{\dagger} 
  c_{-({\bf k}+{\bf q}),-\sigma}^{\dagger}
  c_{-{\bf k},-\sigma} 
  c_{{\bf k},\sigma}.
\end{eqnarray}
Here, we have assumed $g_{\bf q}=g_{-{\bf q}}$. Eq.~\eqref{28} describes the 
renormalization of the $\lambda$ dependent Hamiltonian ${\cal H}_{\lambda}$ 
with respect to the cutoff $\lambda$. Next we use \eqref{28}
to derive flow equations for the parameters 
$\Delta_{{\bf k},\lambda}$ and $C_{\lambda}$. For this 
purpose, a factorization with respect to the full Hamiltonian ${\cal H}$ is 
carried out. The final flow equations read 
\begin{eqnarray}
  \frac{d\Delta_{{\bf k},\lambda}}{d\lambda} &=&
  -2 \sum_{\bf q}
  \delta\left(
    \left| 
      \varepsilon_{\bf k} - \varepsilon_{({\bf k}+{\bf q})}
    \right|
    + \omega_{\bf q} - \lambda
  \right)  
  \frac{
    \left| g_{\bf q} \right|^{2}
    \Theta\left[
      \omega_{\bf q} - 
      \left| \varepsilon_{\bf k} - \varepsilon_{({\bf k}+{\bf q})} \right|
    \right]
  }{
    \left|
      \varepsilon_{\bf k} - \varepsilon_{({\bf k}+{\bf q})} 
    \right| + \omega_{\bf q}
  }
  \left\langle
    c_{-({\bf k}+{\bf q}),\downarrow} c_{({\bf k}+{\bf q}),\uparrow}
  \right\rangle,\nonumber\\[-2ex]
  &&   \label{29} \\[1ex]
  \label{30}
  \frac{dC_{\lambda}}{d\lambda} &=&
  \sum_{\bf k} 
  \left\langle
    c_{{\bf k},\uparrow}^{\dagger} c_{-{\bf k},\downarrow}^{\dagger}
  \right\rangle
  \frac{d\Delta_{{\bf k},\lambda}}{d\lambda}.\\[-3ex]
  && \nonumber 
\end{eqnarray}
Note that in contrast to \eqref{28}, Eqs.~\eqref{29} and \eqref{30} are 
differential equations with normal derivatives of $\Delta_{{\bf k},\lambda}$ 
and $C_{\lambda}$. This fact can be explained as follows: As discussed above, 
the difference between the expressions 
$
  \delta {\cal H}_{1}(\lambda, \Delta \lambda) = 
  {\cal H}_{1,(\lambda -\Delta \lambda)} - 
  {\bf P}_{(\lambda -\Delta \lambda)}{\cal H}_{1, \lambda}
$ 
and 
$
  \Delta {\cal H}_{1}(\lambda, \Delta \lambda) =
  {\cal H}_{1,(\lambda -\Delta \lambda)} - 
  {\cal H}_{1, \lambda}
$ 
is given by the quantity 
${\bf Q}_{(\lambda-\Delta\lambda)} {\cal H}_{1,\lambda}$ which consists of all 
matrix elements of ${\cal H}_{1,\lambda}$ between eigenstates of 
${\cal H}_{0,\lambda}$ with energy differences between 
$(\lambda-\Delta\lambda)$ and $\lambda$. We are interested in the new 
Hamiltonian 
$
  {\cal H}_{(\lambda-\Delta\lambda)} = 
  {\bf P}_{(\lambda-\Delta\lambda)} {\cal H}_{(\lambda-\Delta\lambda)}
$
which only contains transition operators between states with energy 
differences smaller than $(\lambda-\Delta\lambda)$. Therefore, all 
renormalization contributions which lead to matrix elements with energy 
differences larger than $(\lambda-\Delta\lambda)$ are not relevant. Thus, we 
obtain differential equations for the parameters of 
${\cal H}_{(\lambda-\Delta\lambda)}$.

Note that the factor $2$ in front of \eqref{29} is due to the sum over 
$\sigma$ in \eqref{28}. The expectation values 
$\langle \dots \rangle$ in \eqref{29} and \eqref{30} are 
formed with the full Hamiltonian ${\cal H}$ and are independent of  
$\lambda$. The flow equations can be easily integrated between the lower 
cutoff $(\lambda\rightarrow 0)$ and the cutoff $\Lambda$ of the original 
model. The result is
\begin{eqnarray}
  &&\nonumber\\[-3ex]
  \label{31}
  \tilde\Delta_{\bf k} &=&
  \Delta_{{\bf k},\Lambda} + 
  2\sum_{\bf q}
  \frac{
    \left| g_{\bf q} \right|^{2}
    \Theta\left[
      \omega_{\bf q} - 
      \left| \varepsilon_{\bf k} - \varepsilon_{({\bf k}+{\bf q})} \right|
    \right]
  }{
    \left| 
      \varepsilon_{\bf k} - \varepsilon_{({\bf k}+{\bf q})} 
    \right| + \omega_{\bf q}
  }
  \left\langle
    c_{-({\bf k}+{\bf q}),\downarrow} c_{({\bf k}+{\bf q}),\uparrow}
  \right\rangle,\\[2ex]
  \label{32}
  \tilde C &=&
  C_{\Lambda} + 
  \sum_{\bf k} 
  \left\langle
    c_{{\bf k},\uparrow}^{\dagger} c_{-{\bf k},\downarrow}^{\dagger}
  \right\rangle
  \left( \tilde\Delta_{\bf k} - \Delta_{{\bf k},\Lambda} \right)
\end{eqnarray}
where a short hand notation for the desired values of 
$\Delta_{{\bf k}, \lambda}$ and $C_{{\bf k}, \lambda}$ at $\lambda=0$ was 
introduced: 
$
  \tilde\Delta_{\bf k} = \Delta_{{\bf k},(\lambda\rightarrow 0)},$ $
  \tilde C = C_{(\lambda\rightarrow 0)}  
$. 
Note that $\tilde\Delta_{\bf k}$ and $\tilde{C}$ only depend on the 
parameters of the original system \eqref{1} and on $\Delta_{{\bf k},\Lambda}$ 
and $C_{\Lambda}$. The initial conditions \eqref{18} for 
$\Delta_{{\bf k},\Lambda}$ and $C_{\Lambda}$ will be used later. For 
$\lambda\rightarrow 0$ the renormalized Hamiltonian 
$
  \tilde{\cal H} = {\cal H}_{(\lambda\rightarrow 0)} 
$
reads
\begin{eqnarray}
  \label{33}
  \tilde{\cal H} &=&
  \sum_{{\bf k},\sigma} \varepsilon_{\bf k} \,
    c_{{\bf k},\sigma}^{\dagger}c_{{\bf k},\sigma}  +
  \sum_{\bf q} \omega_{\bf q} \,
    b_{\bf q}^{\dagger}b_{\bf q} - 
  \sum_{\bf k}
  \left(
    \tilde\Delta_{\bf k} \,
    c_{{\bf k},\uparrow}^{\dagger} c_{-{\bf k},\downarrow}^{\dagger} + 
    \tilde\Delta_{\bf k}^{*} \,
    c_{-{\bf k},\downarrow} c_{{\bf k},\uparrow} 
  \right) + 
  \tilde{C}.
\end{eqnarray}
$\tilde{\cal H}$ can easily be diagonalized by a Bogoliubov 
transformation according to  
\eqref{19} and \eqref{20}
\begin{eqnarray}
  \label{34}
  \tilde{\cal H} &=&
  \sum_{\bf k} \tilde{E}_{\bf k}
  \left(
    \tilde{\alpha}_{\bf k}^{\dagger} \tilde{\alpha}_{\bf k} +
    \tilde{\beta}_{\bf k}^{\dagger} 
    \tilde{\beta}_{\bf k}
  \right) + 
  \sum_{\bf k}
  \left(
    \varepsilon_{\bf k} - \tilde{E}_{\bf k}
  \right) + 
  \sum_{\bf q} \omega_{\bf q} \, b_{\bf q}^{\dagger}b_{\bf q} + \tilde{C}
\end{eqnarray}
where $\tilde{E}_{\bf k} = E_{{\bf k},(\lambda\rightarrow 0)}$, 
$\tilde\alpha_{\bf k} = \alpha_{{\bf k},(\lambda\rightarrow 0)}$, and 
$\tilde\beta_{\bf k} = \beta_{{\bf k},(\lambda\rightarrow 0)}$. 
\bigskip

Finally, we have to determine the expectation values in \eqref{31} and 
\eqref{32}. Since $\tilde{\cal H}$ emerged from the original model ${\cal H}$ 
by an unitary transformation, the free energy can be calculated 
either from ${\cal H}$ or from $\tilde{{\cal H}}$
\begin{eqnarray}
  \label{35}
  F &=& 
  - \frac{1}{\beta}
  \ln {\rm Tr} \,  e^{-\beta{\cal H}}
  \,=\,
  - \frac{1}{\beta}
  \ln {\rm Tr} \, e^{-\beta\tilde{\cal H}}, \\
  &=&
  - \frac{2}{\beta}
  \sum_{{\bf k}^{\prime}} \ln
  \left(
    1 + e^{-\beta\tilde{E}_{{\bf k}^{\prime}}}
  \right) + 
  \frac{1}{\beta}
  \sum_{\bf q}
  \left(
    1 - e^{-\beta\omega_{\bf q}}
  \right) + 
  \sum_{{\bf k}^{\prime}}
  \left(
    \varepsilon_{{\bf k}^{\prime}} - \tilde{E}_{{\bf k}^{\prime}}
  \right) + \tilde{C} \nonumber
\end{eqnarray}
where \eqref{34} was used. The 
required expectation values are found by functional derivative
\begin{eqnarray}
  \label{36}
  \left\langle
    c_{{\bf k},\uparrow}^{\dagger} c_{-{\bf k},\downarrow}^{\dagger}
  \right\rangle
  &=&
  - \frac{\partial F}{\partial\Delta_{{\bf k},\Lambda}}\\
  &=&
  \sum_{{\bf k}^{\prime}}
  \frac{
    1 - 2f(\tilde{E}_{{\bf k}^{\prime}})
  }{
    2\sqrt{
      \varepsilon_{{\bf k}^{\prime}}^{2} + 
      \left| \tilde\Delta_{{\bf k}^{\prime}} \right|^{2}
    }
  }
  \left[
    \tilde{\Delta}_{{\bf k}^{\prime}}^{*}
    \frac{
      \partial \tilde{\Delta}_{{\bf k}^{\prime}}
    }{
      \partial \Delta_{{\bf k},\Lambda}
    }
    + \tilde{\Delta}_{{\bf k}^{\prime}}
    \frac{
      \partial \tilde{\Delta}_{{\bf k}^{\prime}}^{*}
    }{
      \partial \Delta_{{\bf k},\Lambda}
    }
  \right]
  + \frac{ \partial \tilde{C}}{\partial \Delta_{{\bf k},\Lambda}}
  \nonumber \\[2ex]
  &=&
  \frac{
    \tilde{\Delta}_{\bf k}^{*}
    \left[ 1 - 2f(\tilde{E}_{\bf k}) \right]
  }{
    2\sqrt{
      \varepsilon_{\bf k}^{2} + 
      \left| \tilde\Delta_{\bf k} \right|^{2}
    }
  } + {\cal O}
  \left(
    \left[
      \frac{
        \left| g_{\bf q} \right|^{2}
      }{
        \left|
          \varepsilon_{\bf k} - \varepsilon_{({\bf k}+{\bf q})} 
        \right| + \omega_{\bf q}
      }
    \right]^{2}
  \right).\nonumber
\end{eqnarray}
Here, $f(\tilde{E}_{\bf k})$ denotes the Fermi function with 
respect to the energy $\tilde{E}_{\bf k}$. If we neglect 
higher order corrections, Eqs.~\eqref{31} and \eqref{32} can be rewritten
as
\begin{eqnarray}
  \label{37}
  \tilde\Delta_{\bf k} &=&
  \sum_{\bf q}
  \left\{
    \frac{
      2 \left| g_{\bf q} \right|^{2}
      \Theta\left[
        \omega_{\bf q} - 
        \left| \varepsilon_{\bf k} - \varepsilon_{({\bf k}+{\bf q})} \right|
      \right]
    }{
      \left| 
        \varepsilon_{\bf k} - \varepsilon_{({\bf k}+{\bf q})} 
      \right| + \omega_{\bf q}
    }
  \right\}
  \frac{
    \tilde{\Delta}_{{\bf k}+{\bf q}}^{*}
    \left[ 1 - 2f( \tilde{E}_{{\bf k}+{\bf q}} ) \right]
  }{
    2\sqrt{
      \varepsilon_{{\bf k}+{\bf q}}^{2} + 
      \left| \tilde \Delta_{{\bf k}+{\bf q}} \right|^{2}
    }
  },\\[2ex]
  \label{38}
  \tilde C &=&
  \sum_{\bf k} 
  \left| \tilde{\Delta}_{\bf k} \right|^{2}
  \frac{
    1 - 2f( \tilde{E}_{{\bf k}+{\bf q}} ) 
  }{
    2\sqrt{
      \varepsilon_{{\bf k}+{\bf q}}^{2} + 
      \left| \tilde \Delta_{{\bf k}+{\bf q}} \right|^{2}
    }
  }
\end{eqnarray}
where the initial conditions \eqref{18} were used. Note that Eq.~\eqref{37} 
has the form of the usual BCS-gap equation. Thus, the term 
inside the brackets $\{\dots\}$ can be interpreted as the absolute value of 
the effective phonon-induced electron-electron interaction
\begin{eqnarray}
  \label{39}
  V_{{\bf k},{\bf q}} &=& 
  - \frac{
    2 \left| g_{\bf q} \right|^{2}
    \Theta\left[
      \omega_{\bf q} - 
      \left| \varepsilon_{\bf k} - \varepsilon_{({\bf k}+{\bf q})} \right|
    \right]
  }{
    \left| 
      \varepsilon_{\bf k} - \varepsilon_{({\bf k}+{\bf q})} 
    \right| + \omega_{\bf q}
  }
\end{eqnarray}
for the formation of Cooper pairs. 
In contrast to the usual BCS-theory, in the present formalism 
both the attractive electron-electron 
interaction \eqref{39} and the gap 
equation \eqref{37} were derived in one step 
by applying the renormalization procedure to the 
electron-phonon system \eqref{1}.
\bigskip

Let us now compare the induced electron-elctron interaction \eqref{39} 
with Fr\"{o}hlich's result  \cite{Froehlich}
\begin{eqnarray}
  \label{40}
  V_{{\bf k},{\bf q}}^{\mbox{\tiny \rm Fr\"{o}hlich}} &=& 
  \frac{
    2 \left| g_{\bf q} \right|^{2} \omega_{\bf q}
  }{
    \left[
      \varepsilon_{\bf k} - \varepsilon_{({\bf k}+{\bf q})} 
    \right]^{2} - \omega_{\bf q}^{2}
  }.
\end{eqnarray}
Note that \eqref{40} contains a divergency at 
$|\varepsilon_{\bf k} -\varepsilon_{({\bf k+q}})| = \omega_{\bf q}$. 
Furthermore, this interaction becomes repulsive for
$
  \left|
    \varepsilon_{\bf k} - \varepsilon_{({\bf k}+{\bf q})} 
  \right| > \omega_{\bf q}
$. 
Thus, a cutoff function 
$
  \Theta\left[
    \omega_{\bf q} - 
    \left| \varepsilon_{\bf k} - \varepsilon_{({\bf k}+{\bf q})} \right|
  \right]
$
for the electron-electron interaction \eqref{40} is introduced by hand in the 
usual BCS-theory to suppress repulsive contributions to this interaction. In 
contrast our result \eqref{39} has no divergency and is always
attractive. Furthermore, the cutoff function in Eq.~\eqref{39} shows that
the attractive interaction results from particle-hole excitations with 
energies $|\varepsilon_{\bf k} -\varepsilon_{({\bf k+q}})| < \omega_{\bf q}$.
This result  directly follows from the renormalization process.

Recently, Mielke \cite{Mielke} obtained a $\lambda$-dependent   
phonon-induced electron-electron interaction 
\begin{eqnarray}
  \label{41}
  V_{{\bf k},{\bf q},\lambda}^{\mbox{\tiny \rm Mielke}} &=& 
  - \frac{
    2 \left| g_{\bf q} \right|^{2}
  }{
    \left| 
      \varepsilon_{\bf k} - \varepsilon_{({\bf k}+{\bf q})} 
    \right| + \omega_{\bf q}
  } \ 
\Theta(|\varepsilon_{{\bf k} + {\bf q}}- \varepsilon_{{\bf k}}|
+\omega_{\bf q} -\lambda) 
\end{eqnarray}
where in \eqref{41} 
for convenience the $\lambda$-dependence of the electron and the 
phonon energies is suppressed. Apart from the $\lambda$-dependent 
$\Theta$-function, the main difference to our result \eqref{39} is that 
the cutoff function 
$
  \Theta\left[
    \omega_{\bf q} - 
    \left| \varepsilon_{\bf k} - \varepsilon_{({\bf k}+{\bf q})} \right|
  \right]
$ 
is not present in \eqref{41}. This 
difference may result from Mielke's way of performing the similarity 
transformation \cite{Wilson1,Wilson2} of the electron-phonon system \eqref{1}. 
The similarity transformation is based on the introduction of 
continuous unitary transformations and is formulated in terms of differential 
equations for the parameters of the Hamiltonian. Like in   
our approach (see Sec.~\ref{Ren}) also  
the similarity transformation leads to a 
band-diagonal structure of the normalized Hamiltonian
with respect to the eigen representation  
of the unperturbed Hamiltonian. Due to the 
renormalization processes also new couplings occur. Mielke has 
first evaluated the phonon-induced 
electron-electron interaction \eqref{41} by eliminating excitations 
with energies larger than $\lambda$. 
In particular, for an Einstein model with dispersion-less
phonons of frequency $\omega_0$ the interaction 
becomes independent of $\lambda$ if $\lambda$ is chosen less than 
$\omega_0$. Of course, for this case and assuming 
$
|\varepsilon_{{\bf k} + {\bf q}}- \varepsilon_{{\bf k}}| <
\omega_{\bf q} 
$
Mielke's result \eqref{41} and the result \eqref{39}
become the same. Keeping this interaction and neglecting 
at the same time the remaining part of the electron-phonon 
interaction with energies smaller than $\lambda$ a BCS-like gap equation 
was derived by Mielke\cite{Mielke}. The main difference to our result 
\eqref{39} is the absence of the cutoff-function 
$
  \Theta\left[
    \omega_{\bf q} - 
    \left| \varepsilon_{\bf k} - \varepsilon_{({\bf k}+{\bf q})} \right|
  \right]
$, which demonstrates that only particle-hole excitations with energies less 
than $\omega_q$ participate in the 
attractive interactions. However, note that by setting $\lambda = 0 $ 
a finite value of the interaction remains in \eqref{41}. 
This interaction is non-diagonal in the unperturbed 
Hamiltonian as used by Mielke, 
$
  {\cal H}_{0, \lambda}^{\mbox{\tiny Mielke}} = \sum_{{\bf k},\sigma}
  \varepsilon_{{\bf k},\lambda} 
  c_{{\bf k},\sigma}^{\dagger} c_{{\bf k},\sigma} + 
  \sum_{\bf q} \omega_{\bf q} b_{\bf q}^{\dagger}
  b_{\bf q}
$. 
This seems to be a contradiction to the allowed properties for 
operators at $\lambda=0$ which should commute with the unperturbed 
Hamiltonian.

Finally, by use of Wegner's 
flow equation method \cite{Wegner}, Lenz and Wegner 
obtained the following  phonon-induced electron-electron interaction
\begin{eqnarray}
  \label{42}
  V_{{\bf k},{\bf q}}^{\mbox{\tiny \rm Lenz/Wegner}} &=& 
  - \frac{
    2 \left| g_{\bf q} \right|^{2} \omega_{\bf q}
  }{
    \left[
      \varepsilon_{\bf k} - \varepsilon_{({\bf k}+{\bf q})} 
    \right]^{2} + \omega_{\bf q}^{2}
  }
\end{eqnarray}
which is attractive for all ${\bf k}$ and ${\bf q}$ \cite{Lenz}. The result  
\eqref{42} is similar to \eqref{40} if 
$
  \omega_{\bf q} \geq 
  \left|
    \varepsilon_{\bf k} - \varepsilon_{({\bf k}+{\bf q})} 
  \right|
$
is fulfilled. In contrast to our result \eqref{39}, the interaction 
\eqref{42} remains finite even for 
$
  \left|
    \varepsilon_{\bf k} - \varepsilon_{({\bf k}+{\bf q})} 
  \right| > \omega_{\bf q}
$.
Wegner's flow equation method \cite{Wegner} as well as the similarity 
transformation \cite{Wilson1,Wilson2} are based on the introduction of 
continuous unitary transformations. Both methods are 
formulated in terms of differential equations for the 
parameters of the Hamiltonian.  However, they 
differ in  the generator of the continuous unitary transformation. This leads 
to the different results \eqref{41} and 
\eqref{42}. A detailed comparison of the flow equation 
method and the similarity 
transformation can be found in Ref.~\onlinecite{Mielke}.

%%%%%%%%%%%%%%%%%%%%%%%%%%%%%%%%%%%%%%%%%%%%%%%%%%%%%%%%%%%%%%%%%%%%%%%%%%%%%%%
\section{Conclusion}\label{Conc}

In this paper we have applied a recently developed renormalization 
approach \cite{Becker} to the 'classical' problem of interacting electrons and 
phonons. By adding a small field to the 
Hamiltonian, which break the gauge symmetry, 
we directly derive a BCS-like gap equation for 
the coupled electron-phonon system. 
In particular, it is shown that the derived gap 
function results directly from the renormalization process. The effective 
phonon-induced electron-electron interaction is deduced from the gap 
equation. In contrast to the Fr\"{o}hlich interaction 
\cite{Froehlich} no singularities appear in the effective 
interaction. Furthermore, the cutoff function 
which is included in Fr\"{o}hlich's result by hand 
to avoid repulsive contributions 
to the electron-electron interaction follows directly from the renormalization 
procedure. This means that phonon-induced particle-hole excitations only 
contribute to the attractive electron-electron interaction if 
their energies are smaller than the energy of the 
exchanged phonon.

\section*{Acknowledgements}
We would like to acknowledge helpful discussions with K.~Meyer and T.~Sommer. 
This work was supported by the DFG through the research program SFB 463.

%%%%%%%%%%%%%%%%%%%%%%%%%%%%%%%%%%%%%%%%%%%%%%%%%%%%%%%%%%%%%%%%%%%

\end{document}